\documentclass[usenatbib]{mn2e} \usepackage{psfig}

\title[Roche tomography of CVs -- IV.] {Roche tomography of
cataclysmic variables -- IV. Starspots and slingshot prominences on BV
Cen}

\author[C.\,A.\ Watson, D.\ Steeghs, T.\ Shahbaz and V.\,S.\ Dhillon]
{C.\,A.\ Watson,$^1$\thanks{E-mail: c.watson@sheffield.ac.uk} D.\
Steeghs,$^2$ T.\ Shahbaz,$^3$ and V.\,S.\ Dhillon$^1$ \\ $^1$
Department of Physics and Astronomy, University of Sheffield,
Sheffield S3 7RH, UK\\ $^2$ Harvard-Smithsonian Center for
Astrophysics, 60 Garden Street, Cambridge, MA 02318, USA\\ $^3$
Instituto de Astrof\'{i}sica de Canarias, 38200 La Laguna, Tenerife,
Spain\\}

\date{\center{\Large Submitted for publication in the Monthly
Notices of the Royal Astronomical Society \\
\vspace{.5cm} \today}}

\begin{document}
\maketitle

\begin{abstract}

We present Roche tomograms of the G5--G8 IV/V secondary star in the
long-period cataclysmic variable BV Cen reconstructed from MIKE
echelle data taken on the Magellan Clay 6.5-m telescope. The tomograms
show the presence of a number of large, cool starspots on BV Cen for
the first time. In particular, we find a large high-latitude spot
which is deflected from the rotational axis in the same direction as
seen on the K3--K5 IV/V secondary star in the cataclysmic variable AE
Aqr. BV Cen also shows a similar relative paucity of spots at
latitudes between 40--50$^{\circ}$ when compared with AE
Aqr. Furthermore, we find evidence for an increased spot coverage
around longitudes facing the white dwarf which supports models
invoking starspots at the $L_1$ point to explain the low-states
observed in some cataclysmic variables. In total, we estimate that
some 25 per cent of the northern hemisphere of BV Cen is spotted.

We also find evidence for a faint, narrow, transient emission line
with characteristics reminiscent of the peculiar low-velocity emission
features observed in some outbursting dwarf novae. We interpret this
feature as a slingshot prominence from the secondary star and derive a
maximum source size of 75,000 km and a minimum altitude of 160,000 km
above the orbital plane for the prominence.

The {\em entropy landscape} technique was applied to determine the
system parameters of BV Cen. We find $M_1$ = 1.18 $\pm ^{0.28}_{0.16}$
M$_{\odot}$, $M_2$ = 1.05 $\pm ^{0.23}_{0.14}$ M$_{\odot}$ and an
orbital inclination of $i$ = 53$^{\circ} \pm$ 4$^{\circ}$ at an
optimal systemic velocity of $\gamma$ = --22.3 km s$^{-1}$. Finally,
we also report on the previously unknown binarity of the G5IV star HD
220492.

\end{abstract}


\section{Introduction}
\label{sec:intro}

Cataclysmic variables (CVs) are short period binary systems in which a
(typically) late main-sequence star (the secondary) transfers material
via Roche-lobe overflow to a white dwarf primary star. For an
excellent review of CVs, see \cite{warner95}. Although CVs are largely
observed to study the fundamental astrophysical process of accretion,
it is the Roche-lobe filling secondary stars themselves that are key
to our understanding of the origin, evolution and behaviour of this
class of interacting binary.

In particular, the magnetic field of the secondary star is thought to
play a crucial role in the evolution of CVs -- driving CVs to shorter
orbital periods through {\em magnetic braking}
(e.g. \citealt{kraft67}, \citealt{mestel68}, \citealt{spruit83},
\citealt*{rappaport83}).  Furthermore, the transition of the secondary
star to a fully convective state and the supposed shutdown of magnetic
activity that this transition brings is also invoked to explain the
period gap -- the dearth of CVs with orbital periods between $\sim$
2--3 hours. On more immediate timescales, magnetic activity on the
secondary stars are also thought to explain variations in CV orbital
periods, mean brightnesses, mean outburst durations and outburst
shapes (e.g. \citealt{bianchini90}; \citealt{richman94};
\citealt{ak01}).

That the secondary star and its magnetic field should have such a
large and wide-ranging impact on CV properties should come as no
surprise -- the secondary star essentially acts as the fuel reserve
that powers these binaries. Therefore, an understanding of the
magnetic field properties of the secondary stars in CVs (e.g. spot
sizes, distributions and their variation with time) is crucial if we
are to understand the behaviour of these binaries. In addition,
detailed studies of the rapidly rotating secondary stars in CVs can
also provide tests of stellar dynamo theories under extreme
conditions.  For example, questions regarding the impact of tidal
forces on magnetic flux tube emergence (e.g. \citealt{holzwarth03})
and its effect on differential rotation (e.g. \citealt{scharlemann82})
are particularly pertinent (see \citealt*{watson06} for a discussion).

Despite this, until recently there had been little direct
observational evidence for magnetic activity in CVs. \cite*{webb02}
used TiO bands to infer the presence of spots on the secondary star in
SS Cyg and estimated a spot filling factor of 22 per
cent. Unfortunately, this technique does not allow the surfaces of
these stars to be imaged and hence the spot distributions could not be
ascertained.

Most recently, \cite{watson06} used Roche tomography to map the
starspot distribution on a CV secondary (AE Aqr) for the first
time. In \cite{watson06} we estimated that starspots covered
approximately 18 per cent of the northern hemisphere of AE Aqr. The
Roche tomogram of AE Aqr also showed that starspots were found at
almost all latitudes, although there was a relative paucity of
starspots at a latitude of $\sim$40$^{\circ}$. Furthermore, we found
that, in common with Doppler images of single rapidly rotating stars,
AE Aqr also displayed a large high-latitude spot. In this work we
continue our series of papers on Roche tomography (see
\citealt{watson01a}, \citealt{watson03}, \citealt{watson06}) by
mapping starspots on the long period (0.61-d) dwarf nova BV Cen for
the first time. We also report on the serendipitous discovery of the
binarity of HD220492.

\section{Observations and reduction}

Simultaneous spectroscopic and photometric observations were carried
out over 3 nights on 2004 July 8--10. The spectroscopic data were
acquired using the 6.5-m Magellan Clay Telescope and the simultaneous
photometry was carried out using the Carnegie Institution's Henrietta
Swope 1.0-m Telescope.  Both telescopes are situated at the Las
Campanas Observatory in Chile.

\subsection{Spectroscopy}

The spectroscopic observations of BV Cen were carried out using the
dual-beam Magellan Inamori Kyocera Echelle spectrograph (MIKE --
see~\citealt{bernstein03}).  The MIT Lincoln Labs CCD-20 chip with
2046 $\times$ 4096 pixels was used in the blue channel, and the SITe
ST-002A chip, again with 2046 $\times$ 4096 pixels, was used in the
red channel. The standard setup was used, allowing a wavelength
coverage of 3330\AA~-- 5070\AA~in the blue arm and 4460\AA~--
7270\AA~in the red arm, with significant wavelength overlap between
adjacent orders. With a slit width of 0.7 arcsec, a spectral
resolution of around 38,100 ($\sim$7.8 km s$^{-1}$) and 31,500
($\sim$9.5 km s$^{-1}$) was obtained in the blue and red channels,
respectively. The chips were binned 2$\times$2 resulting in a
resolution element of $\sim$ 2.3 binned pixels in the red arm and 2.6
binned pixels in the blue arm for our chosen slit. The spectra were
taken using 400-s exposure times in order to minimise velocity
smearing of the data due to the orbital motion of the secondary
star. Comparison ThAr arc lamp exposures were taken every $\sim$50
minutes for the purposes of wavelength calibration.

With this setup we obtained 63 usable spectra in each arm. Since the
main goal of the Magellan run was to observe AE Aqr, we were
restricted to 3-hour windows each night to observe BV Cen. Over our
allocated 4 nights this would have allowed over 80 per cent of the
orbit of BV Cen to be observed, but unfortunately we lost the final
night due to bad weather. Other than that, the seeing was typically
0.6--0.7 arcsec on the first night and 0.9 arcsec on the next two
nights, with occasional degradation to 1.5 arcsec. The peak
signal-to-noise of the blue spectra ranged from 29 -- 56 (typically
$\sim$50) in the blue arm, and from 38 -- 76 (typically $\sim$65) in
the red arm.  Table~\ref{table:log} gives a journal of the
observations.

It should be noted that when using MIKE it is not possible to change
the slit orientation on the sky. In order to compensate for this, and
reduce atmospheric dispersion across the slit, MIKE is mounted at a
30$^{\circ}$ angle to the Naysmith platform. This means that at zenith
distances greater than $\sim$50$^{\circ}$ dispersion across the slit
becomes significant. To avoid this, all observations of BV Cen were
carried out at low airmasses and no exposure of BV Cen was carried out
for an airmass above 1.39 (zenith distance $>$ 44$^{\circ}$).

\begin{table*}
\caption[]{Log of the spectroscopic observations of BV Cen, the
relevant spectral-type template stars, a telluric-correction star and
the spectrophotometric standards HR 5501 and HR 9087. The first column
gives the object name, columns 2--4 list the UT start Date and the
exposure start and end times, respectively, and columns 5--6 list the
exposure times and number of spectra taken for each object.  The final
column indicates the type of science frame taken and, where
applicable, the measured systemic velocities, $\gamma$, of the
template stars computed from Gaussian fits to their Least Squares
Deconvolved profiles (see Section~\protect\ref{sec:eph}).}
\begin{tabular}{llccccl} \hline
Object & UT Date & UT Start & UT End & $T_{exp}$ (s) & No. spectra &
Comments \\ \hline

HR 5501 & 2004 Jul 08 & 22:40 & 22:45 & 1--7 & 5 & Spectrophotometric
standard \\ BV Cen & 2004 Jul 08 & 23:34 & 02:31 & 400 & 22 & Target
spectra \\ HR 9087 & 2004 Jul 09 & 10:07 & 10:13 & 1--4 & 6 &
Spectrophotometric standard \\ HR 5501 & 2004 Jul 09 & 22:40 & 22:49 &
1.5 & 12 & Spectrophotometric standard \\BV Cen & 2004 Jul 09 & 23:42
& 02:21 & 400 & 20 & Target spectra \\ Gl 863.3 & 2004 Jul 10 & 10:08
& 10:14 & 15--60 & 4 & G5V template; $\gamma$ = +67.28 $\pm$ 0.10 km
s$^{-1}$ \\ HD 221255 & 2004 Jul 10 & 10:25 & 10:42 & 20--100 & 5 &
G6V template; $\gamma$ = --11.55 $\pm$ 0.10 km s$^{-1}$ \\ HD
224287 & 2004 Jul 10 & 10:52 & 11:04 & 60--200 & 5 & G8V template;
$\gamma$ = +44.57 $\pm$ 0.10 km s$^-1$ \\ HR 5501 & 2004 Jul 10 &
22:41 & 22:47 & 1.5--8 & 8 & Spectrophotometric standard \\ BV Cen &
2004 Jul 10 & 23:35 & 02:18 & 400 & 21 & Target spectra \\ HD 217880 &
2004 Jul 11 & 10:20 & 10:27 & 45--80 & 4 & G8IV template; $\gamma$ =
--67.24 $\pm$ 0.10 km s$^{-1}$ \\ HD 220492 & 2004 Jul 11 & 10:32 &
10:42 & 45--120 & 5 & G5IV template; binary \\ BS 8998 & 2004 Jul 11 &
10:47 & 10:52 & 1.5--8 & 6 & Telluric B9V star \\ \hline
\label{table:log}
\end{tabular}
\end{table*}

\subsubsection{Data reduction}
\label{sec:spec_dr}

The data were reduced using the MIKE {\sc redux} IDL pipeline version
1.7. This automatically processes all calibration frames and then
wavelength calibrates, sky subtracts, flux calibrates and optimally
extracts the target frames. The final output consists of 1-d spectra
for the blue and red arms. The typical rms scatter reported for the
wavelength calibration was around 0.002\AA. Since the {\sc redux}
package outputs vacuum wavelengths, whereas the linelists used in the
Least Squares Deconvolution process (see Section~\ref{sec:lsd}) are
measured in air, the wavelengths have been converted to air using the
IAU standard given by \cite{morton91}.

Unfortunately, we experienced some flux calibration problems around
regions of strong emission lines (e.g. H$\alpha$ and H$\beta$), most
likely due to poor tabulation of our flux standard star HR 5501 around
these lines. Since we mask out the strong emission lines during our
study of the donor star absorption lines, this problem does not affect
this work. We also found that we could not resolve (at the data
reduction stage) a small jump in the flux level between the blue and
red arms. We discuss the solution to this latter problem in
Section~\ref{sec:lsd}.

Since the secondary star contributes a variable amount to the total
light of a CV, Roche tomography is forced to use relative line fluxes
during the mapping process (i.e. it is not possible to employ the
usual method of normalising the data). In order to determine these
relative fluxes, the slit losses need to be calibrated. The standard
technique of using a comparison star on the slit is, however, not
possible with MIKE due to the tightly-packed, cross-dispersed format,
and so simultaneous photometry is required (see
Section~\ref{sec:photometry}). We corrected for slit losses  by
dividing each BV Cen spectrum by the ratio of the flux in the spectrum
(after integrating the spectrum over the appropriate photometric filter
response function) to the corresponding photometric flux (see
Section~\ref{sec:photometry}). We should note here that the slit-loss
correction factors were only calculated using the red spectra, but
applied to both arms of the spectroscopic data. A further correction
was applied to the blue arm (to account for both a different slit-loss
factor in the blue compared to the red and also the jump in flux level
between the two arms mentioned earlier) at the Least Squares
Deconvolution stage (see Section~\ref{sec:lsd}).

\subsection{Photometry}
\label{sec:photometry}

Simultaneous photometric observations were carried out using a Harris
V-band filter and the Site3 CCD chip with 2048 $\times$ 3150
pixels. The CCD chip was windowed to a size of 633$\times$601 pixels
(which covered the target and the brightest comparison stars, as well
as the bias strip) resulting in a 4' x 4' window on the sky. Windowing
allowed the readout time to be reduced to 46-s for each 15-s exposure
on BV Cen.

\subsubsection{Data reduction}

The photometry was reduced using standard techniques. Since the master
bias-frame showed no ramp or large scale structures across the chip,
the bias level of the frames was removed by subtracting the median of
the pixels in the overscan regions. Pixel-to-pixel sensitivity
variations were then corrected by dividing the target frames through
by a master twilight flat-field frame. Optimal photometry was
performed using the package {\sc photom} (\citealt*{eaton02}).

\begin{figure}
\psfig{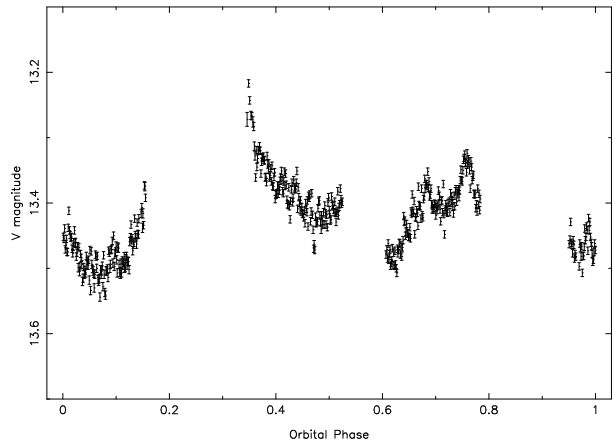}
\caption{The $V$-band light curve of BV Cen. The data have been phased
  according to the ephemeris derived in Section~\protect\ref{sec:eph}. All
3 nights data have been plotted together from phase 0 -- 1.}
\label{fig:light}
\end{figure}

There were two suitable comparison stars on each BV Cen target frame,
identified from the ESO Guide Star Catalogue as GSC0866601471
(hereafter 1471) and GSC0866600859 (hereafter 0859). Two other bright
stars in the field were deemed unsuitable since one was over-exposed,
and the other had a nearby companion. Unfortunately, neither of the
comparisons had reliable magnitude measurements available. We
therefore used observations of the Landolt photometric standard
SA105-437 to calibrate our instrumental magnitudes. This allowed us to
determine the magnitude of 1471 and 0859 as $m_v = 13.43 \pm 0.07$ and
$m_v = 13.54 \pm 0.07$, respectively. Finally, differential photometry
was performed using the bright nearby comparison 1471. The light curve
is shown in Fig.~\ref{fig:light}.

\subsection{Continuum fitting}

The continuum of the BV Cen spectra were fitted in exactly the same
manner as that outlined in the Roche tomography study of AE Aqr
(\citealt{watson06}). Spline knots were placed at locations in the
spectra that were relatively line-free. The spline knot positions were
chosen on an individual spectrum-by-spectrum basis (though the
locations of the knots were generally the same for each spectrum)
until a smooth and visually acceptable fit was obtained. Again, as in
\cite{watson06}, we found that we systematically fit the continuum at
too high a level, leading to continuum regions in the Least Squares
Deconvolved (LSD -- see Section~\ref{sec:lsd}) profiles lying below
zero. The continuum fit was then shifted to lower levels until the
continuum level in the LSD profiles lay at zero. The exact value of
this shift varied from spectrum to spectrum, due to changes in the
continuum shape as a result of variable accretion light, but the
variation was generally small. We found that this process did not
change the actual LSD profile shape.

\section{Least squares deconvolution}
\label{sec:lsd}

Least Squares Deconvolution (LSD) effectively stacks the $\sim$1000's
of photospheric absorption lines observable in a single echelle
spectrum to produce a `mean' profile of greatly increased
signal-to-noise. The technique of LSD is well documented and was first
applied by \cite{donati97a} and has since been used in many Doppler
imaging studies (e.g. \citealt{barnes98}; \citealt{lister99};
\citealt{barnes00}; \citealt{barnes01}; \citealt{jeffers02};
\citealt{barnes04}; \citealt{marsden05}) as well as in the successful
mapping of starspots on AE Aqr (\citealt{watson06}). For detailed
information on LSD, see these references as well as the review by
\cite{cameron01}.

At present we use line lists generated by the Vienna Atomic Line
Database (VALD -- \citealt{kupka99}; \citealt{kupka00}) for LSD. The
spectral type of the secondary star in BV Cen has been determined to
lie within the range G5--G8 IV/V (\citealt{vogt80}).  In our analysis
(see Section~\ref{sec:eph}) later type spectral templates appear to
fit the BV Cen spectra better, with our closest match being a G8IV
template.  In accordance with this, we downloaded a line-list for a
stellar atmosphere with $T_{eff}$ = 5250 K and $\log g = 3.55$ (the
closest approximation available in the database to a G8IV spectral
type).  We do not consider a slight error in the choice of line-list
used as having a significant impact on the results of the LSD process
-- see the discussion on this topic in \cite{watson06} and also
\cite{barnes99}. Since this line-list contained normalised
line-depths, whilst Roche tomography uses continuum-subtracted data,
each line-depth has been multiplied by the corresponding value in a
continuum fit to the G8IV template spectrum.

In order to circumvent the small flux `jump' between the blue and red
arms reported in Section~\ref{sec:spec_dr}, we carried out LSD in each
arm separately. For the red arm, LSD was carried out over a wavelength
range of $\sim$5050\AA~to $\sim$6860\AA, while for the blue arm the
wavelength range was $\sim$4030\AA~to $\sim$5050\AA. Regions of strong
emission lines such as H$\alpha$ and H$\beta$ were also masked
out. Some 2790 lines were used in the deconvolution process across
both arms, leading to a multiplex gain in signal-to-noise of $\sim$22
and yielding absorption profiles with a signal-to-noise of $\sim$1300.
Our version of LSD propagates the errors through the deconvolution
process.

Since the red-spectra were slit-loss corrected using simultaneous
photometry, the resulting LSD profiles from the red data are scaled
correctly relative to one-another. We then rebinned the blue LSD
profiles to the same velocity scale as the red profiles (since the
blue arm had a superior resolution) using sinc-function
interpolation. Slit-loss corrections to the blue LSD profiles were
then made by scaling the blue LSD profiles to match the red profiles
obtained at the same phase. This scaling was done using an optimal
subtraction algorithm. The blue profiles were scaled by a factor $f$
and then subtracted from the red profile. The factor $f$ that resulted
in the minimum scatter in the residuals (compared to a smoothed
version of the residuals) was then used to scale the blue LSD
profiles.  This method allowed the slit-losses in the blue to be
calibrated despite the jump in flux between the red and blue arms. In
addition, this method also allows further loss of light in the blue
due to differential refraction to be corrected to first
order. Deconvolving the blue and red data separately also allowed us
to look for systematic problems and/or differences in the profiles, of
which none were apparent. We should note, however, that the LSD
profiles exhibit a slight tilt, with the red continuum
(more positive velocities) systematically higher than
the continuum on the other side of the profile. This is a well-known
artefact (e.g. \citealt{barnes99}) which will have little impact on
the final image reconstruction.

\begin{figure*}
\psfig{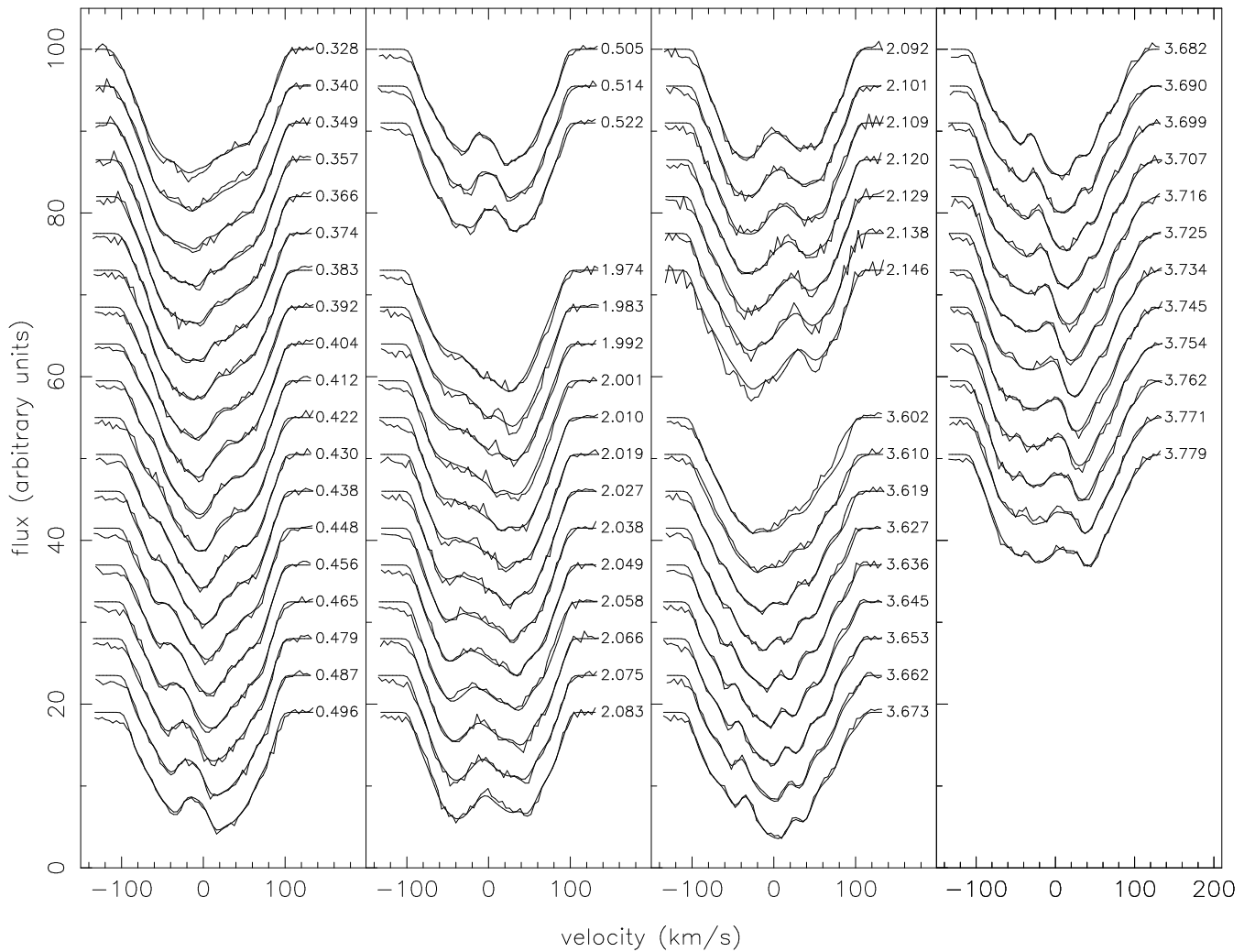}
\caption{The LSD profiles and maximum entropy fits (see
Section\protect~\ref{sec:maps} for further details) for BV Cen. The
velocity of the secondary star has been removed using the parameters
found in Section\protect~\ref{sec:pars}, and each profile is shifted
vertically for clarity. The orbital phase for each exposure is
indicated at the top right of each profile.}
\label{fig:profiles}
\end{figure*}

\begin{figure*}
\hspace{-0.3cm}\psfig{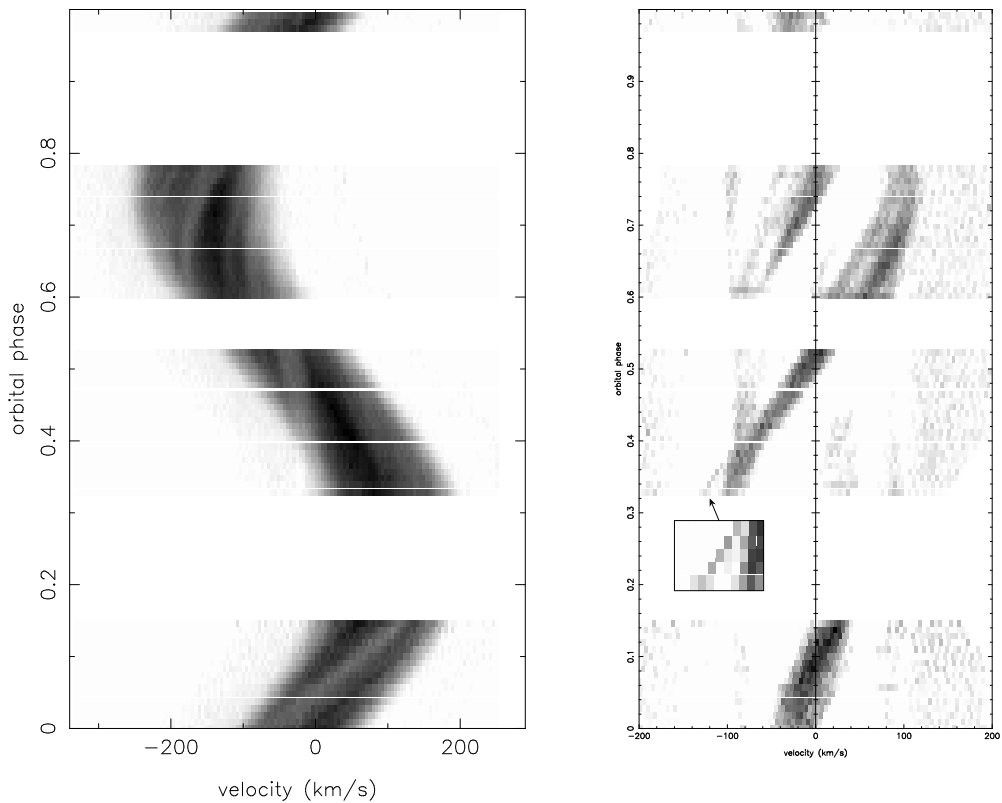}
\caption{Left-hand panel: A trail of the deconvolved profiles of BV
Cen. The large gaps are due to non-continuous observations over three
nights. The small gaps in the phase coverage are at times when arc
spectra were taken for the purpose of wavelength calibration. Features
due to starspots appear bright and several such features are clearly
visible traversing the profiles from blue (more -ve velocities) to red
(more +ve velocities) in the trailed spectra. Also evident is the
orbital motion and variation in $v \sin i$, which shows a maximum at
phase 0.75 due to the varying aspect of the tidally distorted
secondary star. Right-hand panel: The same as the left-hand panel,
except the orbital motion has been removed using the binary parameters
derived in Section~\protect\ref{sec:pars}. This allows the starspot
tracks across the profiles to be more easily followed. In order to
increase the contrast of this plot, we have subtracted a theoretical
profile (see text) from each LSD profile before trailing. Features due
to starspots and irradiation appear dark in these plots. Note the
narrow feature (indicated with an arrow and shown enlarged -- see
inset) that lies outside the blue edge of the profiles at phases
0.328--0.366, which seems to follow the motion of the large feature
that runs through the profiles during this block of observations. This
feature is discussed in more detail in
Section~\protect\ref{sec:sling}.}
\label{fig:trails}
\end{figure*}

The individual LSD profiles are shown in
Fig.~\ref{fig:profiles}. These clearly show the distinct emission
bumps due to starspots moving from blue (-ve velocities) to red (+ve
velocities) through the profile as BV Cen rotates. When the LSD
profiles are trailed (Fig.~\ref{fig:trails}) these features are still
obvious in addition to the secondary stars orbital motion and
variations in the projected equatorial velocity, $v \sin
i$. Fig.~\ref{fig:trails} also shows the trailed LSD profiles after removal
of the orbital motion and subtraction of a theoretical profile at each
phase. The theoretical profiles were calculated using our Roche
tomography code by adopting the binary parameters and limb darkening
described in Section~\ref{sec:pars} and a featureless stellar disc.
This process increases the contrast and enhances the starspot
signatures (which now appear dark).

\section{Roche tomography}
\label{sec:rochey}

Roche tomography is analogous to Doppler Imaging
(e.g.~\citealt{vogt83}) and has now been successfully applied to the
donor stars of CVs on several occasions (\citealt{rutten94};
\citealt{rutten96}; \citealt{watson03}; \citealt{schwope04};
\citealt{watson06}).  Rather than repeat a detailed description of the
methodology and axioms of Roche Tomography here, we refer the reader
to the references above and the technical reviews of Roche Tomography
by \cite{watson01a}, \cite{dhillon01} and \cite{watson04}. The
pertinent points with respect to this work are that we employ a moving
uniform default map, where each element is set to the average value of
the reconstructed map.  Furthermore, we do not adopt a two-temperature
or filling factor model (e.g. \citealt{cameron94}), since these assume
only two temperature components across the star while CV donors are
expected to exhibit large temperature differences due to the impact of
irradiation. This means our Roche tomograms may be prone to the growth
of bright pixels (which two-temperature models suppress,
\citealt{cameron92}) and are quantitatively more difficult to analyse.

\section{Ephemeris}
\label{sec:eph}

\subsection{Ephemeris}
Despite the brightness of BV Cen, and the fact that it is one of the
longest period dwarf novae known, the system has been poorly
studied. The most recent orbital ephemeris for BV Cen was published by
\cite{gilliland82} and will have accumulated a large error over the
last two decades. We have therefore determined a new ephemeris for BV
Cen by cross-correlation with suitable template stars that were
observed using the same instrumental setup. The cross-correlation was
carried out over the spectral range 6400\AA~-- 6536\AA.  This
wavelength range not only covers several moderately strong lines and
blends from the secondary that are useful for radial velocity
measurements but include several temperature and gravity sensitive
lines for F--K stars (\citealt{strassmeier90}).

The BV Cen and template spectra were first rebinned onto the same
velocity scale using sinc-function interpolation and then normalised
by dividing through by a constant. The continuum was then subtracted
by a 3rd order polynomial fit. This procedure is followed to ensure
that the line strengths are preserved across the spectral region of
interest. The template spectra were then broadened to account for the
rotational velocity ($v \sin i$) of the secondary star -- the amount
of broadening applied was determined using an optimal-subtraction
technique. In this method the template spectra were first broadened by
an arbitrary amount before being cross-correlated against the BV Cen
spectra, allowing a first iteration on the radial velocity curve of BV
Cen. The BV Cen data were then orbitally-corrected using the results
of this radial velocity analysis and averaged. The template spectra
were once more broadened by different amounts in 0.1 km s$^{-1}$
steps, multiplied by a constant, $f$, and then subtracted from the
orbitally-corrected mean BV Cen spectrum. The broadening that gave the
minimum scatter in the residual optimally-subtracted spectrum was then
applied to the template spectrum and the whole procedure repeated
until the rotational broadening value no longer changed.  This
typically took 2 or 3 iterations.

Through the above process a cross-correlation function (CCF) was
calculated for each BV Cen spectrum. A radial velocity curve can then
be derived by fitting a sinusoid through the CCF peaks.
Unfortunately, none of our spectral-type templates provided a good
match to the BV Cen spectra, all requiring a multiplication factor $f
> 1$, which would imply that the secondary star contributes more than
100 per cent of the system light. We also discovered that our G5IV
template star HD220492 is actually a close binary (see
Section~\ref{sec:binary}).

\begin{figure}
\psfig{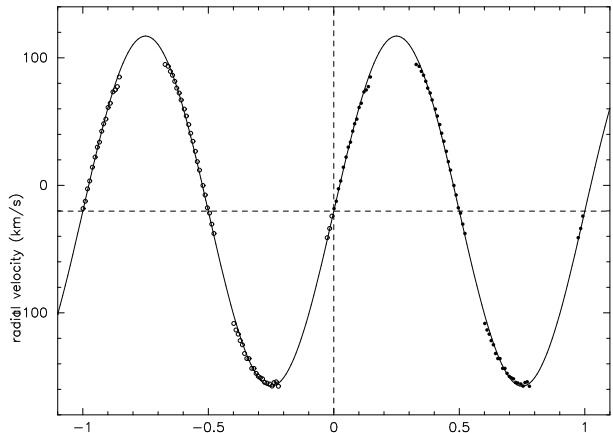}
\caption{The measured secondary star radial-velocities with a
sinusoidal fit from cross-correlation using a G8IV spectral-type
template star. The data have been phase folded using the zero-point of
Section~\protect\ref{sec:eph}. The error bars are typically smaller than
the symbol size.}
\label{fig:xcor}
\end{figure}

Fortunately, as discussed in \cite{watson06}, the CCFs calculated are
insensitive to the use of an ill-matching spectral template and,
despite requiring multiplication factors $f >$ 1, the broadened
template spectra still provided visually acceptable matches to the
orbitally corrected BV Cen spectrum. The radial velocity curve in
Fig.~\ref{fig:xcor} was obtained after cross-correlation of the BV Cen
spectrum with the G8IV star HD 217880 (our best guess spectral type
for BV Cen). This allowed us to obtain a new zero-point for the
ephemeris of
\begin{equation}
T_0 =\, $HJD$\, 2453195.2859 \pm 0.0003
\end{equation}
with the orbital period fixed at $P=0.611179$ d (from
\citealt{gilliland82}). This ephemeris was applied to the rest of the
data presented in this paper.

Cross-correlation with the G8IV template yielded a secondary star
radial velocity amplitude of  $K_r = 137.3 \pm 0.3$ km s$^{-1}$, a
systemic velocity of $\gamma = $ --20.3 $\pm$ 0.2 km s$^{-1}$ and a
rotational broadening $v \sin i$ = 95.3 km s$^{-1}$ (see
Section~\ref{sec:pars} for a discussion of the binary
parameters). These values were consistent within 2 km s$^{-1}$ for all
the spectral-type templates used.  Since some of our spectral-type
templates (including the G8IV template star HD 217880) did not have
measured systemic velocities, we carried out least squares
deconvolution (see Section~\ref{sec:lsd}) on each template star. The
systemic velocity was then measured using a Gaussian fit to the sharp
LSD profile, the results are listed in the final column of
Table~\ref{table:log}. We note that Gl 863.3 has a published systemic
velocity (+66.9 $\pm$ 0.1 km s$^{-1}$; \citealt{nordstroem04}) which
is close to our measured value of +67.28 $\pm$ 0.10 km s$^{-1}$
using this technique.

\subsection{The binarity of HD220492}
\label{sec:binary}

\begin{figure}
\psfig{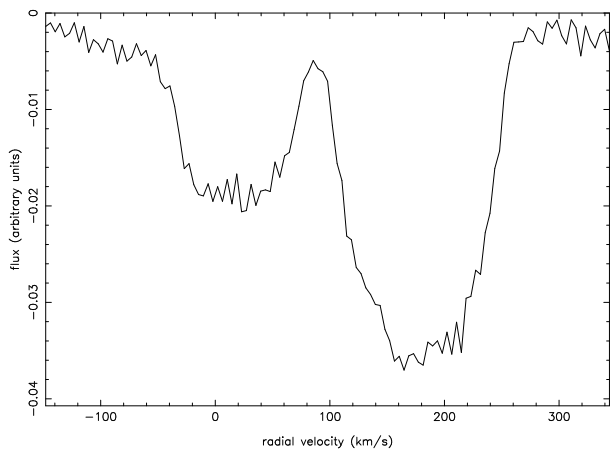}
\caption{The LSD profile of HD220492. This system is most likely a
close binary system containing an early to mid G-type star and a later
type companion.}
\label{fig:hd220492}
\end{figure}

While measuring the systemic velocities of the spectral-type
templates, we discovered that the G5IV spectral-type template HD
220492 is a close binary system. The LSD profile of this system is
shown in Fig.~\ref{fig:hd220492} and clearly shows two heavily
broadened stellar line profiles. It is most likely that this is a
close binary system containing an early to mid G-type star and
later-type companion. We note that the binary nature of HD 220492 was
missed by the Geneva-Copenhagen survey of the Solar neighbourhood,
since no radial velocity measurements of this star appear to have been
taken (\citealt{nordstroem04}). HD 220492 is also listed as the
optical counterpart to the X-ray source 1RXJS232426.0-450906
(\citealt{schwope00a}), and was identified as a suspected variable
star by \cite{samus04}. Given the large $v \sin i$ and hence rapid
rotation of these stars, both components are likely to be magnetically
active. Unfortunately, since only 4 spectra over 7 minutes were taken,
nothing can be said about the orbital period of this system.

\section{system parameters}
\label{sec:pars}

\begin{figure}
\psfig{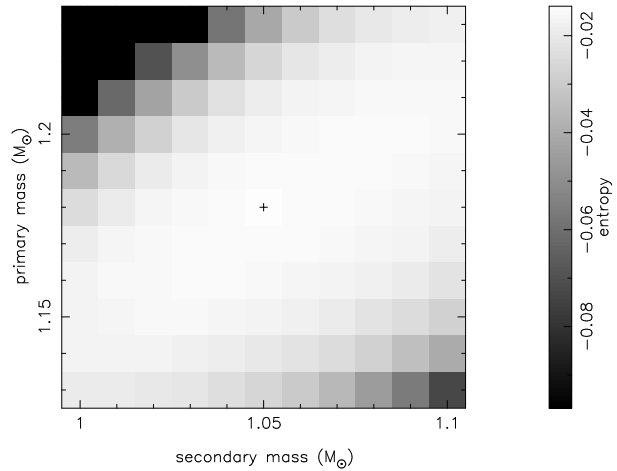}
\caption{The entropy landscape for BV Cen, assuming an orbital
inclination of $i = 53^{\circ}$ and a systemic velocity of $\gamma$ =
--22.3 km s$^{-1}$. Dark regions indicate component masses for which no
fit to the target $\chi^2$ could be found. The cross marks the
position of maximum entropy, corresponding to component masses of
$M_1$ = 1.18 M$_{\odot}$ and $M_2$ = 1.05 M$_{\odot}$.}
\label{fig:land}
\end{figure}

The system parameters (systemic velocity $\gamma$, orbital inclination
$i$, secondary star mass $M_2$ and primary star mass $M_1$) of BV Cen
have been determined using the same method as described in
\cite{watson06}. In short, adopting the incorrect parameters causes
artefacts to appear in the Roche tomograms leading to an increase in
the structure (and hence a decrease in the entropy) of the final
image. This can be visualised as an {\em entropy landscape} (see
Fig.~\ref{fig:land}), in which reconstructions to the same $\chi^2$
are carried out for different combinations of component masses and the
entropy obtained for each set plotted on a grid of $M_1$ versus $M_2$.

In order to search for the correct set of parameters we have
constructed a series of entropy landscapes for different inclinations
and systemic velocities. This is done in an iterative manner by first
constructing entropy landscapes for a sequence of systemic velocities
for a fixed inclination. The optimal systemic velocity found in this
first iteration was then fixed and a series of entropy landscapes were
then calculated out over a range of orbital inclinations.  Once an
optimal orbital inclination was determined another sequence of entropy
landscapes was carried out over a range of systemic velocities to
ensure that the optimal systemic velocity had not changed.  As we have
found previously (\citealt{watson03}; \cite{watson06}), the systemic
velocity obtained in this way is largely independent of the orbital
inclination assumed in the reconstructions.

For the reconstructions of BV Cen we have adopted a root-square
limb-darkening law of the form

\begin{equation}
I\left(\mu\right) =
I_0\left[1-a_{\lambda}\left(1-\mu\right)-b_{\lambda}\left(1-\sqrt{\mu}\right)\right]
\end{equation}
where $\mu = \cos \gamma$  ($\gamma$ is the angle between the line of
sight and the emergent flux), $I_0$ is the monochromatic specific
intensity at the center of the stellar disc, and $a_{\lambda}$ and
$b_{\lambda}$ are the limb-darkening coefficients at wavelength
$\lambda$. We calculated an effective central wavelength of
5067\AA~for our spectroscopic observations using

\begin{equation}
\lambda_{cen} = \frac{\sum_i \frac{1}{\sigma_i^2} d_i \lambda_i}{\sum_i \frac{1}{\sigma_i^2} d_i},
\end{equation}
where $d_i$ and $\sigma_i$ are the line-depths and error on the
observed data at wavelength position $\lambda_i$, respectively. This,
therefore, takes into account the line depths and noise in the
spectrum.  Limb-darkening coefficients for a star of $T_{eff}$ = 5250
K and $\log g$ = 3.5 (corresponding to a G8IV star --
\citealt{dall05}) were obtained from \cite{claret98} for the B- and
V-bands.  The values of the coefficients at these wavelengths  were
then linearly interpolated over to find $a$=0.5388 and $b$=0.3088 for
the effective central wavelength of our observations. These
coefficients were used for all the reconstructions carried out for
this paper.

\begin{figure}
\psfig{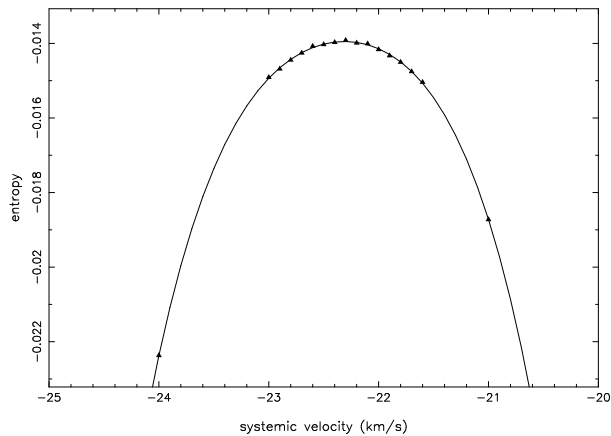}
\caption{Points show the maximum entropy value obtained in each
entropy landscape as a function of systemic velocity, assuming an
orbital inclination of 53$^{\circ}$. The solid curve is a 5th order
polynomial which is only shown to emphasise the trend between
entropy and systemic velocity. The highest data point
corresponds to a systemic velocity of $\gamma$=--22.3 km s$^{-1}$.}
\label{fig:gamma}
\end{figure}

Fig.~\ref{fig:gamma} shows the peak entropy value obtained in entropy
landscapes constructed assuming an orbital inclination of
$i$=53$^{\circ}$ (this value was obtained after carrying out the
iterative procedure described above) but varying the systemic
velocity. This yielded an optimal value of $\gamma$ = --22.3 km
s$^{-1}$. We cannot assign a rigorous error estimate to the systemic
velocity but we found that the image quality depended heavily on the
assumed systemic velocity, and reconstructions were almost impossible
for assumed systemic velocities that differed by more than $\pm$2 km
s$^{-1}$ from the optimal value. Indeed, for this reason we sampled
the systemic velocity more finely than for AE Aqr
(\citealt{watson06}), incrementing in 0.1 km s$^{-1}$ steps. It is
unlikely, therefore, that the error on the systemic velocity
exceeds $\pm$1 km s$^{-1}$ and is probably less than this.

Our value of $\gamma$ = --22.3 km s$^{-1}$, while close to the value
of $\gamma$ = --20.3 $\pm$ 0.2 km s$^{-1}$ derived from the radial
velocity analysis performed in Section~\ref{sec:eph}, is in stark
contrast to the two previous published values known to us. These are
--47$\pm$10 km s$^{-1}$ (\citealt{vogt80}) and --47$\pm$2 km s$^{-1}$
(\citealt{gilliland82}). We are uncertain as to the cause of this
discrepancy. It is unlikely that surface inhomogeneities, such as
those caused by irradiation from the accretion regions (and which is
known to cause systematic errors in radial velocity measurements --
e.g. \citealt{davey92}; \citealt{watson03}), could cause this
discrepancy. First, our systemic velocity derived from radial velocity
analysis only differs from that derived from the entropy landscape
method by 2 km s$^{-1}$, which implies that inhomogeneities on the
surface of BV Cen have little impact. Second, one could argue that
during the observations of \cite{vogt80} and \cite{gilliland82} the
impact of irradiation may have been much greater. In the case of HU
Aqr, which is heavily irradiated, we found that the systemic velocity
measured by standard radial velocity analyses could be shifted by as
much as --14 km s$^{-1}$ from the true value
(\citealt{watson03}). Such a shift, if applicable to BV Cen, would
bring the results of \cite{gilliland82} and \cite{vogt80} into closer
agreement with ours but only if a large portion of the leading
hemisphere (the side of the star as seen from $\phi$ = 0.75) was
irradiated during their observations. The lightcurve of \cite{vogt80}
does indicate that the system was slightly brighter during their
observations, but the system was certainly not in outburst and we
would not expect a large amount of irradiation of the donor
star. Furthermore, looking at our Roche tomograms in
Section~\ref{sec:maps}, irradiation of BV Cen appears to be fairly
low, and certainly not large enough to impact a radial velocity curve
analysis to the extent needed to bring our values of the systemic
velocity into agreement. Nor do we believe that our wavelength
calibration is incorrect, as we have measured the systemic velocity of
Gl 863.3 using LSD and found it to be in agreement with
\cite{nordstroem04} to within 0.4 km s$^{-1}$ (see
Section~\ref{sec:eph}). Therefore, we have no satisfactory explanation
of the discrepancy between our measured systemic velocity and those of
\cite{vogt80} and \cite{gilliland82}.

\begin{figure}
\psfig{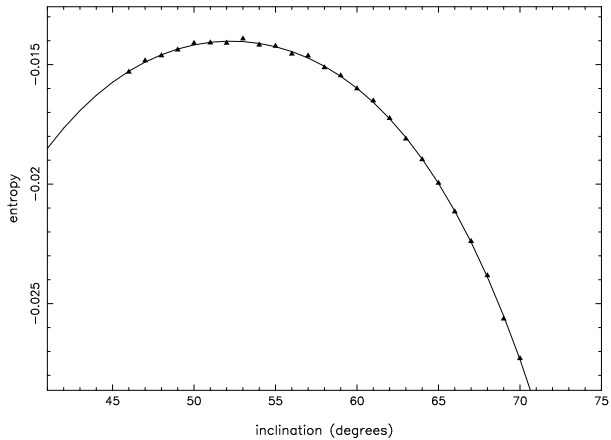}
\caption{Points show the maximum entropy value obtained in each
entropy landscape for different inclinations, assuming a systemic
velocity of $\gamma$ = --22.3 km s$^{-1}$. The solid curve is a fifth
order polynomial fit showing the general trend. The entropy value
peaks at an orbital inclination of $i$=53$^{\circ}$. Inclinations
lower than 49$^{\circ}$ result in an optimal white dwarf mass greater
than the Chandrasekhar limit and so we have not been extended the
analysis to significantly lower inclinations.}
\label{fig:incl}
\end{figure}

Fig.~\ref{fig:incl} shows the maximum entropy values as a function of
the orbital inclination assuming the systemic velocity of $\gamma$ =
--22.3 km s$^{-1}$ derived earlier. This shows a clear trend with the
best fitting inclination at $i$ = 53$^{\circ}$ which we have adopted
as the optimal value for BV Cen. This value is lower than (but agrees
to within 2-$\sigma$ with) the orbital inclinations of 62$^{\circ}\pm$
5$^{\circ}$ and 61$^{\circ}\pm$ 5$^{\circ}$ of \cite{gilliland82} and
\cite{vogt80}, respectively.

The entropy landscape for BV Cen carried out with $i$ = 53$^{\circ}$
and $\gamma$ = --22.3 km s$^{-1}$ is shown in
Fig.~\ref{fig:land}. From this we derive optimal masses for the white
dwarf and secondary star of $M_1$ = 1.18 M$_{\odot}$ and $M_2$ = 1.05
M$_{\odot}$, respectively.  These values for the binary parameters
seem to fall in between the previously published ones. \cite{vogt80}
also find relatively high masses for the binary components of $M_1$ =
1.4 $\pm$ 0.2 and $M_2$ = 1.4 $\pm$ 0.2. Indeed, their derived white
dwarf mass places it uncomfortably close to the Chandrasekhar limit
while our new estimate lies more comfortably within the allowed mass
limit for white dwarfs.  On the other-hand, \cite{gilliland82} find
lower masses of $M_1$ = 0.83 $\pm$ 0.1 and $M_2$ = 0.9 $\pm$ 0.1. These
masses, however, lead to a high mass ratio ($q$ = $M_2$/$M_1$) of $q =
1.08 \pm 0.18$, which places BV Cen near the critical mass ratio for
mass transfer instability (\citealt{politano96};
\citealt{thoroughgood04}).  Our higher secondary star mass, and lower
mass ratio ($q = 0.89$) moves BV Cen comfortably within the region
where mass transfer is stable. We should note that, since the optimal
inclination we find is quite low, a small error in the derived
inclination leads to quite a large error in the masses. As expected,
we find that the optimal masses in our entropy landscapes vary as
$\sin^3 i$.

Selection of the correct $\chi^2$ to aim for during the
reconstructions is somewhat subjective. We select the $\chi^2$ which
results in a close fit to the data (all reconstructions were carried
out to $\chi^2$=2.5, which indicates that our propagated errors are
underestimated but this does not impact the final reconstructions),
but not so close as to cause the Roche tomograms to break up into
noise. In order to determine how robust the derived parameters are, we
have explored the effects that changing the aim $\chi^2$ has on the
entropy landscapes. We find that fitting more closely to the data
causes the maps to start to break up into noise, but we find that the
derived inclination is the same ($i$ = 53$^{\circ}$), and the masses
for each entropy landscape are typically within $\pm$0.02 M$_{\odot}$,
and we never see a difference greater than $\pm$0.04 M$_{\odot}$ for
any assumed inclination. The same is true if we raise the aim $\chi^2$
(poorer fit to the data), except we lose sensitivity to the
inclination. Despite this, even fitting to substantially higher
$\chi^2$'s where the maps become featureless we still find a
preference for inclinations between $\sim$52 -- 60$^{\circ}$.  We are
therefore confident in the robustness of our derived binary parameters.

Calculating the errors on our derived parameters is difficult. One
could perform a series of Monte Carlo simulations of the data and
carry out entropy landscape reconstructions for each synthetic
dataset. Unfortunately, this would require a large amount of time to
complete. Instead we can determine a rough guide to the errors from
the scatter in the vsini and $K_r$ values obtained from conventional
analysis of the data using the spectral-type templates described in
Section~\ref{sec:eph}. We found that the scatter in the measured $K_r$
and vsini values were, in both cases, $\pm$ 2 km s$^{-1}$. This leads
to an error on the mass ratio, $q$, of $\pm$ 0.04 -- which in turn
translates into an error on the component masses of $\pm$0.03
M$_{\odot}$. Such an error agrees with the scatter we find in the
optimal values returned from entropy landscapes constructed for
different $\chi^2$'s. This then gives optimal parameters of $M_1$ =
1.18 $\pm$ 0.03 M$_{\odot}$ and $M_2$ = 1.05 $\pm$ 0.03 M$_{\odot}$ at
an inclination of 53$^{\circ}$.

Clearly, the errors in the masses are dominated by any error in the
derived orbital inclination.  Again, we can make a conservative
estimate to the likely error in our orbital inclination. A lower limit
of $i$ = 49$^{\circ}$ is set by the neccesity that the white dwarf
mass be lower than the Chandrasekhar limit.  If we assume that,
therefore, our inclination is accurate to $i$=53$\pm$4$^{\circ}$, we
find $M_1$ = 1.18 $\pm ^{0.28}_{0.16}$ M$_{\odot}$ and $M_2$ = 1.05
$\pm ^{0.23}_{0.14}$ M$_{\odot}$.

\section{surface maps}
\label{sec:maps}

\begin{figure}
\psfig{figure=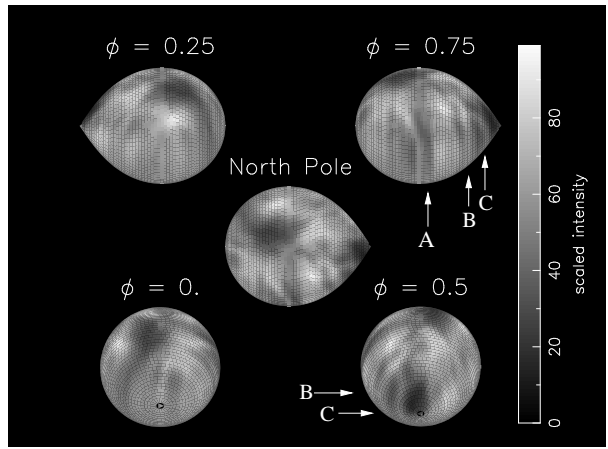,width=8.0cm}
\caption{The Roche tomogram of BV Cen. Dark grey-scales indicate the
presence of either star-spots or irradiated zones. The system has been
plotted as the observer would see it at an orbital inclination of
$53^{\circ}$, except for the central panel which shows the system as
viewed from above the North Pole. The orbital phase (with respect to
the ephemeris of Section~\ref{sec:eph}) is indicated above each panel.
For clarity, the Roche tomograms are shown excluding limb darkening.
Individual spots referred to in the text are arrowed.}
\label{fig:bvcen}
\end{figure}

Using the parameters derived in Section~\ref{sec:pars} we have
constructed a Roche tomogram of the secondary star in BV Cen (see
Fig.~\ref{fig:bvcen}). The corresponding fit to the data are displayed
in Fig.~\ref{fig:profiles}. A number of dark starspots are visible  in
the Roche tomogram. The most conspicuous of these is the high latitude
spot (common in Doppler images of rapidly rotating stars) at a
latitude of $\sim$65$^{\circ}$. Such high-latitude spots seen on
rapidly rotating stars are commonly believed to be caused by strong
Coriolis forces acting to drive magnetic flux tubes towards the poles
(\citealt{schussler92}).  This spot is also similar to the large
high-latitude spot seen on AE Aqr by \cite{watson06}, not just in its
latitude but also in the way that it appears towards the trailing
hemisphere of the star. If such a shift in high latitude spots on CV
donor stars is common then this hints that a mechanism is in
operation that drives flux-tube emergence to this side of the star. We
discuss this in more detail later.

Also prominent is a dark feature near the $L_1$ point. Although it is
possible that this may be due to a large spot (with obvious
consequences for theories that invoke starspots to quench mass
transfer), we have to consider that the secondary stars in CVs are
located close to an irradiating source (the bright spot and/or white
dwarf, for example). It is well known that the inner face of the
secondary, and the $L_1$ point in particular, is often
irradiated. Indeed, early single-line Roche tomography studies of CVs
concentrated on mapping the so-called irradiation patterns on the
inner faces of the secondaries in CVs (e.g. \citealt{watson03}). While
it is largely the contrast between the immaculate photosphere and
lower spotted continuum contributions that cause spots to appear dark
on our tomograms, irradiation causes the absorption lines to weaken
due to ionisation and hence also appears dark.

We believe that the feature near the $L_1$ point in BV Cen is indeed
due to irradiation. This interpretation is supported by the modelling
of B--V lightcurves by \cite{vogt80} who found evidence for a slightly
enhanced temperature around the inner Lagrangian point.
\cite{gilliland82} also found evidence for a slightly heated inner
face.  Close inspection shows that the irradiation pattern is
asymmetric, with the impact of irradiation strongest towards the
leading hemisphere. This is very similar to the irradiation pattern
seen on the dwarf-nova IP Peg (see \citealt{watson03} and
\citealt{davey92}) which has been explained by irradiation from the
bright spot which is located on the correct side of the star to create
this asymmetry. Certainly the lightcurves of BV Cen
(e.g. \citealt{vogt80}) show a prominent bright spot component. Also
\cite{vogt80} suggested that the B--V variations could be explained if
a small region of the star nearest to the gas stream and bright spot
had a temperature excess of $\Delta T \sim$ 300 K. The location of
this heated region described by \cite{vogt80} matches what we observe
in the Roche tomogram.

Several other features in the Roche tomogram are also worth
mentioning. One of these is a mid-latitude spot best seen at phase
$\phi$ = 0.75 (marked as spot A on Fig~\ref{fig:bvcen}), which we have
mentioned as it is one of the largest spots, other than the polar
spot. Interestingly, there also seem to be two low-latitude spots near
the $L_1$ point (spots 'B' and 'C' on Fig~\ref{fig:bvcen}), but
located on the leading hemisphere. Although these spots are fairly
heavily smeared in latitude, they may be located at a sufficiently low
latitude to cross the mass-transfer nozzle. It has been suggested that
starspots are thought to be able to quench mass transfer as they pass
the mass-losing nozzle on the donor, resulting in the low-states
observed in many CVs (see \citealt{livio94}; \citealt{king98}).  The
observations of low-latitude starspots near the $L_1$ point lends some
credence to these models -- it seems apparent that sizeable starspots,
which are often seen at high latitudes on rapidly-rotating stars, can
also form at low latitudes near the $L_1$ point in binaries.

It is unlikely that the spots we have identified near the L$_1$ point
are, instead, due to irradiation. This would imply that the
irradiation is patchy in nature (rather than smoothly varying across
the stellar surface) in order to form the dark spots we observe in the
tomogram. This, in turn, would require that the irradiation is either
beamed towards relatively small patches on the stellar surface, or
would require a complex accretion structure shielding the majority of
the stellar surface and only allowing small spot-like regions to be
irradiated. We can think of no reason why the irradiation should be
beamed in this manner, nor why such a complex accretion structure
should exist. In addition, while the limb-darkening around the $L_1$
region could be incorrect by a significant degree due to the low
effective-gravity, again this would be a smoothly varying artifact and
would not lead to dark spots on the Roche tomogram.

\begin{figure}
\psfig{figure=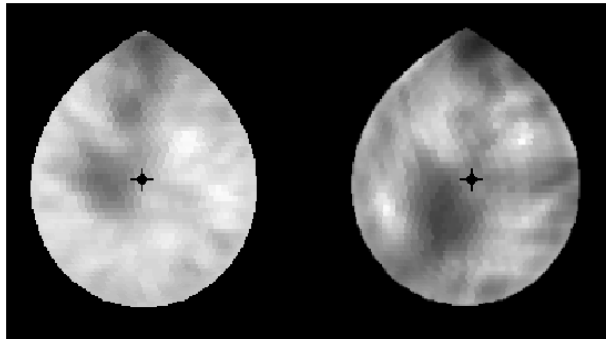,width=8.0cm}
\caption{Roche tomograms of the donor stars in AE Aqr (left -- from
\citealt{watson06}) and BV Cen (right) as viewed from above the pole
(indicated with a cross). Both show a high latitude spot displaced
towards the left-most (trailing) hemisphere (the orbital motion is
towards the right). Furthermore, both appear to show a chain of spots
descending from the polar spot towards the $L_1$ point.}
\label{fig:plan}
\end{figure}

We also note that, when viewed from above the North pole of the star,
there seems to be quite a distinct `chain' of spots. This chain leads
down from the polar regions towards the $L_1$ point, showing a
possible deflection towards the leading hemisphere with decreasing
latitude. In Fig.~\ref{fig:plan} we plot an enlarged polar view of
both BV Cen and AE Aqr (from \citealt{watson06}) which shows more
clearly these features. Comparison with the Roche tomogram of AE Aqr
shows a similar distribution of spots (we make a more detailed
comparison in Section~\ref{sec:discussion}). The fact that we see this
same `chain' of spots is suggestive of a mechanism that is forcing
magnetic flux tubes to preferentially arise at these locations. Given
that  this is on the side of the star facing the white dwarf, this may
be due to the impact of tidal forces which is thought to be able to
force spots to arise at preferred longitudes (\citealt{holzwarth03}).
In a recent Doppler Image of the pre-CV V471 Tau, \cite{hussain06}
also found the presence of high-latitude spots located on the side of
the star facing the white dwarf.

In addition to the features outlined above, a number of other small
spots are visible. We have examined the intensity distribution of the
pixels on our Roche tomogram in order to make a more quantitative
analysis of the spot distribution. In our Roche tomography study of AE
Aqr (\citealt{watson06}), we looked for a bimodal distribution in
pixel intensities and labelled the population of lower intensity
pixels as spots, and higher intensity pixels as immaculate
photosphere.  Unfortunately, unlike AE Aqr, BV Cen does not show a
clear bimodal distribution in pixel intensities from which we can
confidently distinguish between immaculate and spotted photosphere
(note that we neglected regions that are not visible and therefore
have pixel intensities set to the default value). Instead, the
histogram of pixel intensities shows a broad peak, with long tails
towards high and low pixel intensities. We have therefore defined a
spot intensity by examining the polar spot feature. At the centre of
the polar spot, the pixel intensities are very stable, and we have
taken the intensity of the central regions of the polar spot to
represent the intensity of a 100\% spotted region. Although there are
lower intensity pixels present in the Roche tomogram (which contribute
to the tail of low intensity pixels in the maps), all of these were
found to be confined to the irradiated zone near the $L_1$ point. (We
should note that the fact that the region around the $L_1$ point has
pixel intensities quite different from those of the spotted regions
supports our interpretation that this feature is not a spot). We are
therefore reasonably confident in our identification of the spotted
regions on BV Cen.

Determining the immaculate photosphere was somewhat more
difficult. There appear to be some bright regions in the map,
especially near the polar spot.  It is unlikely that these represent
the immaculate photosphere; the growth of bright pixels in maps that
are not `thresholded' is a known artifact of Doppler imaging methods
(e.g. \cite{hatzes92} -- also see Section~\ref{sec:rochey}), and it is
likely that these features are due to this, contributing to the long
tail of bright pixel intensities that we see. Instead, we have
selected the intensity at the upper end of the broad peak of pixel
intensities in the histogram to represent the immaculate
photosphere. If we assume that the appearance and disappearance of
spots simply alters the continuum level (and not the line-depths,
which has a secondary influence on the LSD profiles) and a blackbody
scaling, this gives a temperature contrast between the photosphere and
spot of $\Delta T$ = 780K. Such a temperature difference seems
reasonable. From simultaneous modelling of lightcurves and line-depth
ratios of three active RS CVn systems, \cite{frasca05} found a
temperature difference between spotted and immaculate photosphere of
$\Delta T$ = 450--850 K. Similarly, \cite{biazzo06} find temperature
differences of $\Delta T$ = 453--1012K for stars at different
locations along the HR diagram, with stars of lower surface gravity
having spots with a lower $\Delta T$.

Each pixel in our Roche tomogram was given a spot-filling factor
between 0 (immaculate) to 1 (totally spotted) depending on its
intensity between our predefined immaculate and spotted photosphere
intensities. After removal of the region near the $L_1$ point (which
is caused by irradiation and would cause us to overestimate the total
spot coverage), we find that some 25\% of the visible hemisphere of BV
Cen is spotted. This figure is relatively high compared to the spot
coverages returned from Doppler images of isolated stars (typically 10
per cent) and is probably due to our less than ideal characterisation
of spot filling-factors compared to imaging codes that invoke
two-temperature models. However, a visual comparison of BV Cen with
the map of AE Aqr (\citealt{watson06} -- for which we estimated had 
18\% spot coverage) clearly shows that BV Cen has a higher coverage of
spots on its surface, and our estimated spot coverage of 25\% for BV Cen
is consistent with this. Furthermore, a TiO study by \cite{webb02} found
a 22\% spot coverage for the CV SS Cyg, comparable to what we find for
BV Cen.

\begin{figure*}
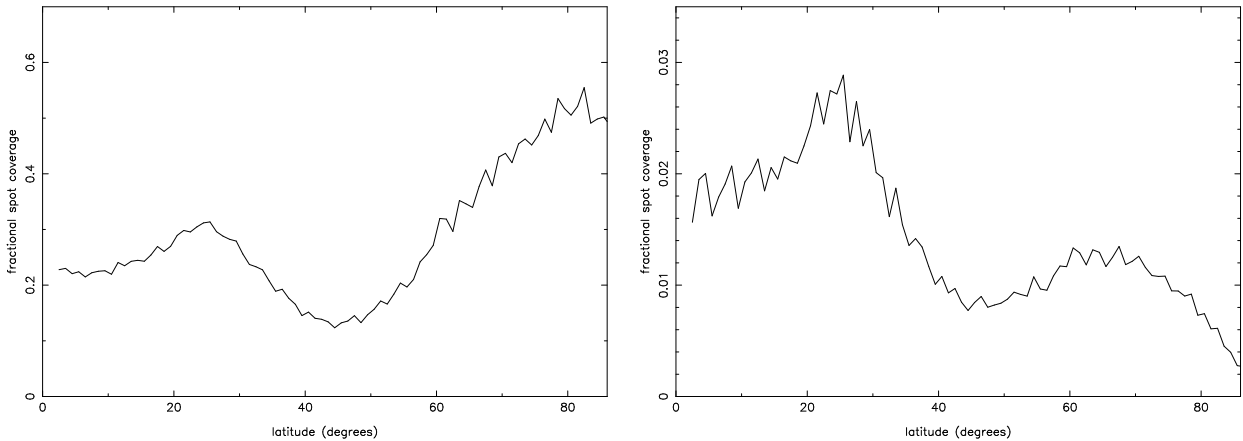

\begin{tabular}{cc}
\psfig{figure=spotfraction2.ps,width=8.0cm,angle=-90.} &
\psfig{figure=spotfraction1.ps,width=8.0cm,angle=-90.} \\
\end{tabular}
\caption{Plots showing the spot coverage on BV Cen as a function
of latitude. Left: spot coverage as a function of latitude expressed in
terms of the surface area at that latitude. Right: spot coverage as a
function of latitude normalised by the total surface area of the
northern hemisphere. Since the grid in our Roche tomograms is not
aligned along strips of constant latitude, some interpolation between
grid elements is required in order to produce these plots. This
results in their slightly noisy nature.}
\label{fig:spotfraction}
\end{figure*}

Based on our spot characterisation scheme outlined above, we have
determined the distribution of starspots as a function of latitude
scaled either by the total surface area of the star, or by the surface
area of the latitude strip over which the spot coverage was calculated
(see Fig.~\ref{fig:spotfraction}). This shows that the highest
spot-filling factor is achieved at polar latitudes, primarily due to
the large high-latitude spot centred at +65$^{\circ}$. We also find that
the spot coverage reaches a minimum at intermediate latitudes around
$\sim$45$^{\circ}$. This is also very similar to what was found for AE Aqr
(\citealt{watson06}), though the drop in spot coverage at this latitude
is far more evident in BV Cen than for AE Aqr. 

Fig.~\ref{fig:spotfraction} also shows clear evidence for a bimodal
distribution of starspots with latitude. Our analysis indicates a
lower-latitude site of spot emergence around 25$^{\circ}$; indeed the
majority of spots seem to form at lower latitudes in BV Cen. In our
Roche tomography study of AE Aqr (\citealt{watson06}), we found some
evidence for spots at lower latitudes, although this was somewhat
uncertain given the low signal-to-noise of those observations. While
the BV Cen data has large phase gaps, this time we can be confident
that we are seeing spots at lower latitudes. Many of the features
described here, such as the off-pole high latitude spot, the bimodal
spot distribution and the relative paucity of spots at intermediate
latitudes are also seen in Doppler Images of a number of other stars
(see the discussion in \citealt{watson06} and references therein).
What is most startling, however, is the great similarity between
the maps of AE Aqr presented in \cite{watson06}, and those of BV Cen
presented in this work.

\begin{figure*}
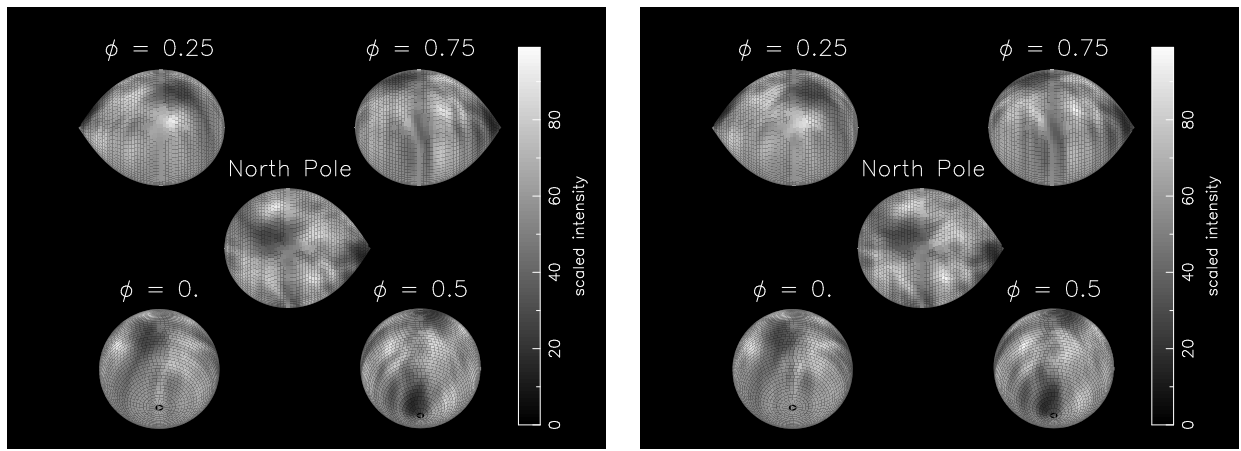

\begin{tabular}{ll}
\psfig{figure=bvcen_odd.ps,width=8.0cm,angle=-90.} &
\psfig{figure=bvcen_even.ps,width=8.0cm,angle=-90.} \\
\end{tabular}
\caption{As for Fig.~\protect\ref{fig:bvcen}, but reconstructions for
odd-numbered spectra (left) and even-numbered spectra (right).}
\label{fig:map2}
\end{figure*}

Finally, we have checked whether the features seen in the Roche tomogram
are real, or artefacts due to noise. We carried out two further Roche
tomography reconstructions, the first only using the odd numbered spectra,
and the other using the even numbered spectra. These are both shown in
Fig.~\ref{fig:map2} and, despite the reduction in phase coverage, both
maps show the same features. We are therefore confident that the
features in the Roche tomogram are not due to noise.

\section{A slingshot prominence?}
\label{sec:sling}

In addition to the many starspot signatures that are visible in BV
Cen's trailed spectra (see Fig~\ref{fig:trails}), a curious narrow
feature is also evident between phases 0.328 and 0.366 on the blue
edge of the profile. This feature appears as a narrow continuation of
the main track through the profiles during this block of observations,
but lies outside the stellar absorption profile and therefore cannot
lie on the stellar surface. Closer inspection of the individual
profiles (see Fig.~\ref{fig:sling}) reveals a narrow, weak emission
feature in  the continuum of 5 LSD profiles. In order to confirm the
reality of this feature, we have visually inspected the profiles that
were deconvolved from the blue and red spectra separately to see
whether they are present in both sets of profiles. Indeed this feature
is present (albeit very weakly) in both sets of profiles which makes
it unlikely to be due to noise or a systematic effect arising during
the LSD process. This feature is also not due to contamination from
lunar or solar light during the observations as this would appear as
an absorption feature (e.g. \citealt{marsden05}). Furthermore,
moon-rise did not occur until at least 2 hours (extending to 4 hours
on the final night) after observations of BV Cen were concluded.  The
fact that the feature is at zero velocity with respect to BV Cen's
systemic velocity of -22 km s$^{-1}$ rules out a `terrestrial' origin,
which would be centred at 0 km s$^{-1}$. We are therefore confident
that this feature is real.

Since the emission feature lies outside the stellar line profile, and
hence lies off the stellar limb, it can only be attributed to
circumstellar material. Solar prominences appear as bright emission
loops when they are viewed off the solar limb, and it is probable that
we are also seeing a prominence structure on BV Cen. Indeed, large
prominences have been reported on other rapidly rotating stars such as
AB Dor (e.g. \citealt{cameron89}) and Speedy Mic
(\citealt{dunstone06a}) and are often observed as transient absorption
features passing through the Doppler broadened H$\alpha$ stellar
line. Recently, however, analysis of VLT data of Speedy Mic by
\citet{dunstone06b} has revealed rotationally modulated emission
outside of the stellar H$\alpha$ line due to loops of emission seen
off of the stellar disc, but which can also be associated with
prominences seen to transit the stellar disc at other times. In
addition, peculiar low-velocity emission features seen in SS Cyg and
IP Peg during outburst have also been interpreted as `slingshot
prominences' (\citealt{steeghs96}). \cite{gansicke98} discovered the
presence of highly-ionized low velocity-dispersion material located
between the $L_1$ point and the centre of mass in AM Her which they
attributed to a slingshot prominence. Similarly, triple-peaked
H$\alpha$ lines following the motion of the donor star in AM Her has
been reported by \cite{kafka05} and \cite*{kafka06}. The authors
interpreted these as long-lived prominences on the donor star and
noted that one component was consistent with stellar activity lying
vertically above the $L_1$ point.

The emission feature we see in BV Cen appears stationary at $\sim$0 km
s$^{-1}$ within the binary frame. This is in keeping with the stationary
slingshot prominences seen in SS Cyg and IP Peg by \cite{steeghs96},
and the low-velocity emission observed in other CVs
(\citealt{marsh90b}). Generally, we would expect the emission from
prominences observed off the stellar limb to be weak and undetectable
for CVs given their faintness.  However, it is possible that the
prominences are illuminated by the accretion light, causing
prominences forming between the donor star and the white dwarf to
become visible when otherwise they would be undetectable. Certainly,
the low velocity of the emission suggests a position close to the
centre-of-mass of the binary at a point between the donor star and
white dwarf where such illumination is most likely.

Given that prominences are normally only seen in lines that form above
the photosphere (e.g. the Hydrogen Balmer lines), observing them in
the LSD profiles which are obtained from photospheric lines is
unexpected under normal conditions. We believe that the emission seen
in BV Cen's photospheric lines is due to excitation of these species
within low density gas in the prominence due to the impact of
irradiation. Thus, we suspect that not only does irradiation cause the
prominence to become highly visible, but also causes it to be
observable in some photospheric lines in which prominences in a
normal, unirradiated environment would not normally emit. We have
considered that the emission could be due to a wind launched from the
accretion regions. Such a wind, however, would exhibit a
radial velocity modulation due to the orbital motion of the primary
star which we do not observe. Furthermore, presumably the velocities
of material in such a wind would produce a far broader emission line
than observed in BV Cen. For these reasons, and the fact that
it kinematically matches previous observations of slingshot
prominences in CVs, we prefer the interpretation that this feature is
the result of an irradiated prominence.

\subsection{Limits on the prominence size and height}

We find that the emission feature in BV Cen is very narrow, with a
velocity width ($\Delta V$) of $\sim$ 10 km s$^{-1}$. Using this width
we can place an upper limit on the emission source size, $l$. Following
\cite{steeghs96}, we assume the prominence is co-rotating with the
secondary star, so we can write,
\begin{equation}
\frac{\Delta V}{K_1 +K_2} = \frac{l}{a},
\end{equation}
where $K_1$ and $K_2$ are the radial velocity amplitudes of the white
dwarf and donor star, respectively, and $a$ is the orbital separation.
For the parameters derived for BV Cen in Section~\ref{sec:pars}, this
places an upper limit of 75,000 km. Naturally, the velocity dispersion
is increased by instrumental resolution, thermal broadening,
saturation broadening and turbulence within the prominence itself. For
a 10,000 K prominence the thermal Doppler velocity of Hydrogen is 12.9
km s$^{-1}$ (e.g. \citealt{dunstone06b}). Since the lines included in
our LSD are mainly heavier elements such as Fe and Ca, the thermal
broadening will be far smaller at $\sim$2 km s$^{-1}$. Solar
prominences exhibit turbulent motions of several km s$^{-1}$, and
\cite{dunstone06b} estimate a turbulent velocity of 5 km s$^{-1}$ for
the prominences observed on Speedy Mic. However, the dominant
broadening mechanism is the $\sim$9.5 km s$^{-1}$ instrumental
resolution of our observations.  Considering these limitations, our
estimated maximum source size of 75,000 km should be viewed with
caution as it is possible that the source is actually unresolved.

We have further analysed the LSD profiles to check whether or not we
can see the zero-velocity prominence as it tracks across the stellar
disc. We were unable to positively identify any feature with
zero-velocity between phases 0.374--0.522 during the first night's
data.  This may be due to the large emission bump feature that
traverses the profile due to irradiation and/or starspots (see
Section~\ref{sec:maps}) which probably masks the prominence
itself. Analysis of the second night's data, however, does reveal a
narrow emission bump at zero-velocity between orbital phases
1.974--2.058. This is quite conspicuous (see Fig.~\ref{fig:sling}) as
the emission feature moves across the profile in the opposite
direction to the starspot features, and Roche Tomography is clearly
unable to fit it. This agrees with the picture of a prominence holding
material near the centre-of-mass of the binary, and that at orbital
phase $\sim$0 we are effectively looking over the top (or possibly
under the bottom) of the star at the prominence on the other side.
This, therefore, means that the prominence structure must be raised
above or below the orbital plane in order for it not to be eclipsed by
the donor star at this phase.

Given the lack of eclipse and the parameters found for BV Cen in
Section~\ref{sec:pars}, combined with the assumption that the
prominence is located above the centre-of-mass of the binary (the
point of zero-velocity), we find that the prominence must lie at least
160,000 km above the orbital plane. If we assume that the prominence
lies below the centre-of-mass (i.e. we are looking `underneath' the
star around phase 0) then it lies out of the orbital plane by at least
2,400,000 km. We feel that this latter case is highly unlikely. First,
a prominence below the orbital plane is far more likely to be eclipsed
by the accretion regions around orbital phase $\sim$0.35, when it is
clearly seen in the data. Second, if we assume that the prominence is
only visible due to illumination from the accretion regions then a
prominence below the orbital plane is located too far away from the
irradiating source.

\subsection{Prominence evolution and structure}

\begin{figure}
\hspace{0.5cm}\psfig{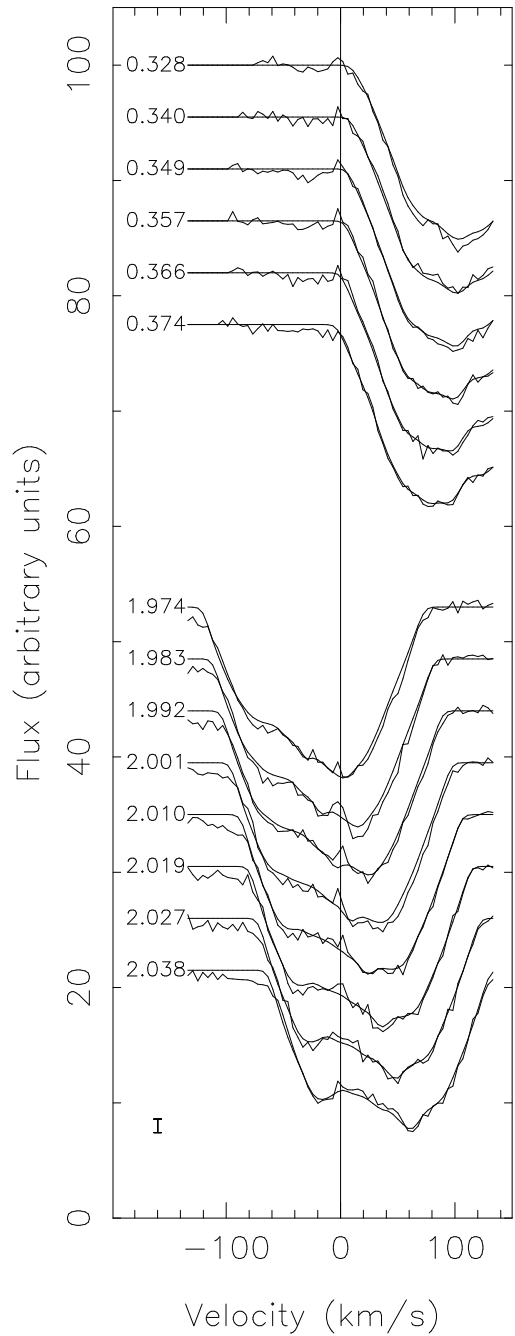}
\caption{Plots of the LSD profiles that show the narrow emission
feature at a velocity of $\sim$0 km s$^{-1}$ on nights 1 \& 2 as
discussed in the text. The Roche tomography fits to the data are also
plotted and the orbital phase is indicated on the left. Unlike
Fig.~\protect\ref{fig:profiles}, the orbital motion has not been
removed and a vertical line centred on a radial velocity of 0 km
s$^{-1}$ has been plotted to show where the emission feature appears.
The slingshot prominence feature appears outside of the stellar lines
on the first night's data (top 6 profiles). During the second night
(bottom profiles) it appears within the profiles but moves in the
opposite direction relative to the starspot features and the Roche
tomography code is therefore unable to fit this feature.}
\label{fig:sling}
\end{figure}

Three nights of consecutive (albeit interrupted) observations allows a
limited discussion of the evolution of the slingshot prominence
observed on BV Cen. The prominence is certainly seen at the start of
the first night ($\phi$ = 0.328--0.366) before it transits the stellar
disc, where it then becomes invisible. We have assumed that, rather
than the prominence disappearing at this point, its signature is lost
in the complex structure present in the stellar line profile. We are
able to pick up an emission feature with the same position in velocity
space again at the start of the second night ($\phi$ = 1.974--2.038,
see Fig.~\ref{fig:sling}).

Curiously, this prominence feature then disappears after orbital phase
2.038. Since, unlike the first night, the prominence on the second
night appears quite clearly in the middle of the stellar line profile
it is, on this occasion, difficult to explain how the feature could
suddenly be lost within the stellar line. This suggests that the
disappearance may be due to rapid evolution of the prominence
itself. Indeed, it does seem as though the emission feature weakens
during the second night before its apparent disappearance (see
Fig.~\ref{fig:sling}). Though the interpretation of a prominence
rapidly evolving on timescales of hours is speculative,
\cite{dunstone06b} also find evidence of individual prominences
evolving on timescales of $\sim$9 hours on Speedy Mic. Certainly, we
can find no evidence for the prominence feature on the 3rd night. This
is despite the fact that the emission should be well separated from
the stellar line profile after orbital phase 3.645. This supports the
idea that the prominence material is not long-lived, only lasting a
couple of days before either draining back to the stellar surface or
being ejected. \cite{dunstone06a} found that, while some prominences
on Speedy Mic were still visible after 5 nights, others formed or
disappeared over the course of one night.

Finally, unlike the prominences seen on Speedy Mic and AB Dor which
lie typically between 2 to 9 stellar radii from the stellar rotation
axis (with a concentration at the co-rotation radius), the feature
observed in BV Cen is far closer to the stellar surface. If the
prominence is located above the centre-of-mass of the binary, then
this places it just 1.5 R$_*$ (assuming the Roche-lobe volume radius
for BV Cen) from the rotation axis of the secondary star. Although
this is at odds with most H$\alpha$ observations which show
prominences at or beyond the co-rotation radius, clouds substantially
closer to the stellar surface have been reported in HK Aqr
(\citealt*{byrne96} reported prominences 0.34 -- 3.2 R$_*$ above the
stellar surface) and RE J1816+542 (heights as low as 0.88 R$_*$,
\citealt{eibe98}).

The height of the prominence feature seen in BV Cen also  agrees with
X-ray observations of rapidly-rotating isolated stars, even when
H$\alpha$ observations of prominences of the stars in question reveal
high cloud heights. For instance, in Chandra X-ray observations of AB
Dor, \citet{hussain05} found that a significant fraction of the
emission arose from compact regions near the stellar surface with
heights of less than 0.3 R$_*$, and that the emitting corona does not
extend more than 0.75 $R_*$ above the stellar surface.  Obviously, the
formation mechanism of these prominences and active regions
necessitates more work. In particular, the apparent preference for
`slingshot prominences' observed in CVs to form at low-velocity sites
requires satisfactory explanation.

\section{Discussion}
\label{sec:discussion}

We have shown that the donor star in BV Cen is highly spotted, the
second CV donor for which this has been shown to be the case.
Fig.~\ref{fig:plan} highlights the striking similarities in spot
distribution between BV Cen and AE Aqr, despite their quite different
binary parameters and spectral types. Although BV Cen has a higher
spot coverage and more low-latitude spots than AE Aqr, both show a
high latitude spot displaced towards the trailing hemisphere. It would
be interesting if all donor stars in CVs showed high latitude spots
displaced in the same direction. Such a deflection could be the result
of the orbital motion of the binary. Since the rotation axis lies
outside the donor stars in these binaries, one may expect the play of
Coriolis (and/or centrifugal) forces on magnetic flux tube emergence
to be different from single stars.  Further observations, however, are
required before we can definitely say that such a deflection of
high-latitude spots takes place in CV donors.

Another striking similarity between the two tomograms presented in
Fig.~\ref{fig:plan} is the apparent chain of spots extending down from the
polar regions to the $L_1$ point. This is more evident in the tomogram
of AE Aqr, but probably only because there are fewer low latitude
features which allows these spots to stand-out more on AE Aqr.  We
believe that this chain of spots is probably due to the impact of
tidal forces due to the close proximity of a compact companion, and
such a `sub-white dwarf' concentration of spots is also seen on the
pre-CV V471 Tau (\citealt{hussain06}). A concentration of starspots on
the inner face of the donor stars in CVs may also have consequences
for the accretion dynamics of these objects. It has long been thought
that starspots may be able to quench mass-transfer from the donor as
they pass across the $L_1$ point, leading to the low-states seen in
many CVs (e.g. \citealt{livio94}, \citealt{king98}). Indeed, in their
study of the mass-transfer history of AM Her, \cite*{hessman00}
concluded that such a model would require an unusually high
spot-coverage near the $L_1$ point, or otherwise some mechanism that
drives spots towards the $L_1$ point. Certainly, both AE Aqr and BV
Cen do seem to show increased spot coverages towards the $L_1$ point in
support of their conclusions.

The fact that we see more active regions near the $L_1$ point may also
explain why we appear to see a preference for `slingshot prominences'
to form above the donors inner face (\citealt{steeghs96};
\citealt{gansicke98}; \citealt{kafka05}; \citealt{kafka06} -- and
again in this work). Certainly, the fact that prominence material at
these locations will be illuminated by the accretion regions also
means that observations will be biased towards detecting prominences
in the region between the white dwarf and donor star in
CVs. Furthermore, if surface magnetic fields are strong enough in the
neighbourhood of the $L_1$ point this may also cause fragmentation of
the mass flow, resulting in the inhomogeneous or `blobby' accretion
seen in some CVs (e.g. \citealt{meintjes04}; \citealt{meintjes06}).

\section{conclusions}
\label{sec:conclusions}

BV Cen is the second CV donor star for which starspots have
unambiguously been imaged on. Again, as with AE Aqr, we find a high
(25 per cent) spot coverage, and the high activity-level is further
confirmed by the detection of a slingshot prominence. Comparison with
the Roche tomograms of AE Aqr from \cite{watson06} show many
similarities between the two systems despite the quite different
fundamental parameters of these two binaries. These spot distributions
hint at the impact of tidal and/or Coriolis forces on the emergence of
magnetic flux tubes in these binaries, and suggest that the inner
faces of CV donor stars are unusually heavily spotted. As always,
further observations are recommended in order to confirm that such
spot distributions are widely seen on CV donors. If a fixed spot
distribution were to be found, this would also suggest that
differential rotation is suppressed (or at least weak) in CV donors,
as suggested by \cite{scharlemann82}.

\section*{\sc Acknowledgements}

CAW is supported by a PPARC Postdoctoral Fellowship. DS acknowledges
a Smithsonian Astrophysical Observatory Clay Fellowship as well as
support through NASA GO grant NNG06GC05G. TS acknowledges
support from the Spanish Ministry of Science and Technology under the
programme Ram\'{o}n y Cajal. The authors acknowledge the use of the
computational facilities at Sheffield provided by the Starlink
Project, which is run by CCLRC on behalf of PPARC. The Starlink
package {\sc photom} was used in this work. We would like
to thank the Observatories of the Carnegie Institution of Washington
for generously allowing us the use of the Henrietta Swope telescopes
at Las Campanas Observatory. The 6.5m Landon Clay (Magellan II) telescope
at Las Campanas is operated by the Magellan consortium consisting of
the Carnegie Institution of Washington, Harvard University, MIT,
the University of Michigan, and the University of Arizona.

\bibliographystyle{mn2e}

\end{document}